\documentclass{emulateapj}

\usepackage{listings}
\usepackage{amsmath}

\newcommand{\culsp}{\textsc{culsp}}
\newcommand{\lsp}{\textsc{lsp}}

\newcommand{\diff}{\ensuremath{{\rm d}}}

\newcommand{\Pnorm}{\ensuremath{P_{\rm n}}}
\newcommand{\Pnormc}{\ensuremath{P_{\rm n}^{\culsp}}}
\newcommand{\Pnorml}{\ensuremath{P_{\rm n}^{\lsp}}}

\newcommand{\freq}{\ensuremath{f}}
\newcommand{\dfreq}{\ensuremath{\Delta f}}
\newcommand{\freqNy}{\ensuremath{f_{\rm Ny}}}
 
\newcommand{\Nt}{\ensuremath{N_{t}}}
\newcommand{\Nfreq}{\ensuremath{N_{\freq}}}
\newcommand{\Nfreqind}{\ensuremath{N_{\freq, {\rm ind}}}}
\newcommand{\Nblock}{\ensuremath{N_{\rm b}}}

\newcommand{\ct}{\ensuremath{c_{\tau}}}
\newcommand{\st}{\ensuremath{s_{\tau}}}
\newcommand{\XS}{\ensuremath{X\!S}}
\newcommand{\XC}{\ensuremath{X\!C}}
\newcommand{\CC}{\ensuremath{CC}}
\newcommand{\SSa}{\ensuremath{SS}}
\newcommand{\CS}{\ensuremath{CS}}

\newcommand{\tcalc}{\ensuremath{t_{\rm calc}}}
\newcommand{\tcalcm}{\ensuremath{\langle\tcalc\rangle}}
\newcommand{\tcalcs}{\ensuremath{\sigma(\tcalc)}}

\newcommand{\Fover}{\ensuremath{F_{\rm over}}}
\newcommand{\Fhigh}{\ensuremath{F_{\rm high}}}

\newcommand{\VAql}{V1449~Aql}

\begin{document}


\title{Fast Calculation of the Lomb-Scargle Periodogram Using Graphics
  Processing Units} 
\shorttitle{Fast Calculation of the Lomb-Scargle
  Periodogram Using Graphics Processing Units}


 \author{R. H. D. Townsend}\affil{Department of Astronomy, University
   of Wisconsin-Madison, Sterling Hall, 475 N. Charter Street,
   Madison, WI 53706, USA; townsend@astro.wisc.edu}

\shortauthors{Townsend}


\begin{abstract}
  I introduce a new code for fast calculation of the Lomb-Scargle
  periodogram, that leverages the computing power of graphics
  processing units (GPUs). After establishing a background to the
  newly emergent field of GPU computing, I discuss the code design and
  narrate key parts of its source. Benchmarking calculations indicate
  no significant differences in accuracy compared to an equivalent
  CPU-based code. However, the differences in performance are
  pronounced; running on a low-end GPU, the code can match 8 CPU
  cores, and on a high-end GPU it is faster by a factor approaching
  thirty.  Applications of the code include analysis of long
  photometric time series obtained by ongoing satellite missions and
  upcoming ground-based monitoring facilities; and Monte-Carlo
  simulation of periodogram statistical properties.
\end{abstract}



\section{Introduction} \label{sec:intro}

Astronomical time-series observations are often characterized by
uneven temporal sampling (e.g., due to transformation to the
heliocentric frame) and/or non-uniform coverage (e.g., from day/night
cycles, or radiation belt passages). This complicates the search for
periodic signals, as a fast Fourier transform (FFT) algorithm cannot
be employed. A variety of alternatives have been put forward, the most
oft-used being the eponymous Lomb-Scargle (L-S) periodogram developed
by \citet{Lom1976} and \citet{Sca1982}. At the time of writing, NASA's
Astrophysics Data System (ADS) lists 735 and 1,810 publications
(respectively) that cite these two papers, highlighting how important
the L-S periodogram has proven for the analysis of time series. Recent
applications include the search for a link between solar rotation and
nuclear decay rates \citep{Stu2010}; the study of pulsar timing noise
\citep{Lyn2010}; the characterization of quasi-periodic oscillations
in blazars \citep{Ran2010}; and the measurement of rotation periods in
exoplanet host stars \citep{Sim2010}.

Unfortunately, a drawback of the L-S periodogram is a computational
cost scaling as $\mathcal{O}(\Nt^{2})$, where \Nt\ is the number of
measurements in the time series; this contrasts with the
far-more-efficient $\mathcal{O}(\Nt \log_{2} \Nt)$ scaling of the FFT
algorithm popularized by \citet{CooTuk1965}. One approach to reducing
this cost has been proposed by \citet{PreRyb1989}, based on
constructing a uniformly sampled approximation to the observations via
`extirpolation' and then evaluating its FFT. The present paper
introduces a different approach, not through algorithmic development
but rather by leveraging the computing power of graphics processing
units (GPUs) --- the specialized hardware at the heart of the display
subsystem in personal computers and workstations. Modern GPUs
typically comprise a number of identical programmable processors, and
in recent years there has been significant interest in applying these
parallel-computing resources to problems across a breadth of
scientific disciplines. In the following section, I give a brief
history of the newly emergent field of GPU computing; then,
Section~\ref{sec:l-s} reviews the formalism defining the L-S
periodogram, and Section~\ref{sec:culsp} presents a GPU-based code
implementing this formalism. Benchmarking calculations to evaluate the
accuracy and performance of the code are presented in
Section~\ref{sec:calc}. The findings and future outlook are then
discussed in Section~\ref{sec:discuss}.

\section{Background to GPU Computing} \label{sec:back}

\subsection{Pre-2006: Initial Forays} \label{ssec:back-initial}

The past decade has seen remarkable increases in the ability of
computers to render complex 3-dimensional scenes at movie
frame-rates. These gains have been achieved by progressively shifting
the graphics pipeline --- the algorithmic sequence of steps that
converts a scene description into an image --- from the CPU to
dedicated hardware within the GPU. To address the inflexibility that
can accompany such hardware acceleration, GPU vendors introduced
so-called \emph{programmable shaders}, processing units that apply a
simple sequence of transformations to input elements such as image
pixels and mesh vertices. NVIDIA Corporation were the first to
implement programmable shader functionality, with their GeForce 3
series of GPUs (released March 2001) offering one vertex shader and
four (parallel) pixel shaders. The release in the following year of
ATI Corporation's R300 series brought not only an increase in the
number of shaders (up to 4 vertex and 8 pixel), but also capabilities
such as floating-point arithmetic and looping constructs that laid the
foundations for what ultimately would become GPU computing.

Shaders are programmed using a variety of specialized languages, such
as the OpenGL Shading Language \citep[GLSL; e.g.,][]{Ros2006} and
Microsoft's High-Level Shading Language (HLSL). The designs of these
languages are strongly tied to their graphics-related purpose, and
thus early attempts at GPU computing using programmable shaders had to
map each calculation into a sequence of equivalent graphical
operations \citep[see, e.g.,][and references therein]{Owe2005}. In an
effort to overcome this awkward aspect, \citet{Buc2004} developed
BrookGPU --- a compiler and run-time implementation of the Brook
stream programming language for GPU platforms. With BrookGPU, the
computational resources of shaders are accessed through a \emph{stream
  processing} paradigm: a well-defined series of operations (the
\emph{kernel}) are applied to each element in a typically-large
homogeneous sequence of data (the \emph{stream}).

\subsection{Post-2006: Modern Era} \label{ssec:back-modern}

GPU computing entered its modern era in 2006, with the release of
NVIDIA's \emph{Compute Unified Device Architecture} (CUDA) --- a
framework for defining and managing GPU computations without the need
to map them into graphical operations. CUDA-enabled devices \citep[see
Appendix A of][]{NVI2010} are distinguished by their general-purpose
unified shaders, which replace the function-specific shaders (pixel,
vertex, etc.) present in earlier GPUs. These shaders are programmed
using an extension to the C language, which follows the same
stream-processing paradigm pioneered by BrookGPU. Since the launch of
CUDA, other vendors have been quick to develop their own GPU computing
offerings, most notably Advanced Micro Devices (AMD) with their
\emph{Stream} framework, and Microsoft with their \emph{DirectCompute}
interface.

Abstracting away the graphical roots of GPUs has made them accessible
to a very broad audience, and GPU-based computations are now being
undertaken in fields as diverse as molecular biology, medical imaging,
geophysics, fluid dynamics, economics and cryptography
\citep[see][]{Pha2005,Ngu2007}. Within astronomy and astrophysics,
recent applications include $N$-body simulations \citep{Bel2008},
real-time radio correlation \citep{Way2009}, gravitational lensing
\citep{Tho2010}, adaptive-mesh hydrodynamics \citep{Sch2010} and
cosmological reionization \citep{AubTey2010}.


\section{The Lomb-Scargle Periodogram} \label{sec:l-s}

This section reviews the formalism defining the Lomb-Scargle
periodogram. For a time series comprising \Nt\ measurements $X_{j} \equiv
X(t_{j})$ sampled at times $t_{j}$ ($j=1,\ldots,\Nt$), \emph{assumed
  throughout to have been scaled and shifted such that its mean is
  zero and its variance is unity}, the normalized L-S periodogram at
frequency \freq\ is
\begin{multline} \label{eqn:Pnorm}
\Pnorm(\freq) = \frac{1}{2} \left\{ 
  \frac{\left[\sum_{j} X_{j} \cos \omega(t_{j} - \tau)
    \right]^{2}}
       {\sum_{j} \cos^{2} \omega(t_{j} - \tau)} + \mbox{}
     \right. \\
\left. \frac{\left[\sum_{j} X_{j} \sin \omega(t_{j} - \tau)
  \right]^{2}}
     {\sum_{j} \sin^{2} \omega(t_{j} - \tau)}
\right\}.
\end{multline}
Here and throughout, $\omega \equiv 2\pi\freq$ is the angular
frequency and all summations run from $j=1$ to $j=\Nt$. The
frequency-dependent time offset $\tau$ is evaluated at each $\omega$
via
\begin{equation} \label{eqn:tau}
\tan 2 \omega \tau = \frac{\sum_{j} \sin 2\omega t_{j}}{\sum_{j}
  \cos 2\omega t_{j}}.
\end{equation}
As discussed by \citet{Sch1998}, \Pnorm\ in the case of a pure
\emph{Gaussian}-noise time series is drawn from a beta
distribution. For a periodogram comprising \Nfreq\
frequencies\footnote{The issue of `independent' frequencies is briefly
  discussed in Section~\ref{ssec:discuss-app}.}, the \emph{false-alarm
  probability} (FAP) --- that some observed peak occurs due to chance
fluctuations --- is
\begin{equation} \label{eqn:fap}
Q = 1 - \left[1 - \left( 1 - \frac{2\Pnorm}{\Nt} \right)^{(\Nt-3)/2} \right]^{\Nfreq}.
\end{equation}

Equations~(\ref{eqn:Pnorm}) and~(\ref{eqn:tau}) can be written
schematically as
\begin{equation}
\Pnorm(\freq) = \sum_{j} \mathcal{G}[\freq, (t_{j}, X_{j})],
\end{equation}
where $\mathcal{G}$ is some function. In the classification scheme
introduced by \citet{Bar2010}, this follows the form of an
\emph{interact algorithm}. Generally speaking, such algorithms are
well-suited to GPU implementation, since they are able to achieve a
high arithmetic intensity. However, a straightforward implementation
of equations~(\ref{eqn:Pnorm}) and~(\ref{eqn:tau}) involves two
complete runs through the time series to calculate a single
$\Pnorm(\freq)$, which is wasteful of memory bandwidth and requires
$\Nfreq (4 \Nt + 1)$ costly trigonometric function evaluations for the
full periodogram. \citet{Pre1992} address this inefficiency by
calculating the trig functions from recursion relations, but this
approach is difficult to map onto stream processing concepts, and
moreover becomes inaccurate in the limit of large \Nfreq. An
alternative strategy, which avoids these difficulties while still
offering improved performance, comes from refactoring the equations as
\begin{multline} \label{eqn:Pnorm-refac}
\Pnorm(\freq) = \frac{1}{2} \left[ 
  \frac{(\ct \XC + \st \XS)^{2}}
       {\ct^{2} \CC + 2 \ct\st \CS + \st^{2} \SSa} + \mbox{} 
\right. \\
\left. 
  \frac{(\ct \XS - \st \XC)^{2}}
       {\ct^{2} \SSa - 2 \ct \st \CS + \st^{2} \CC}
\right],
\end{multline}
and
\begin{equation} \label{eqn:tau-refac}
\tan 2 \omega \tau = \frac{2\,\CS}{\CC-\SSa}.
\end{equation}
Here, 
\begin{equation}
 \ct = \cos \omega \tau, \qquad \st = \sin \omega \tau,
\end{equation}
while the sums
\begin{align} \nonumber
\XC & = \sum_{j} X_{j} \cos\omega t_{j}, \\ \nonumber
\XS & = \sum_{j} X_{j} \sin\omega t_{j}, \\ \label{eqn:sums} 
\CC & = \sum_{j} \cos^{2}\omega t_{j}, \\ \nonumber
\SSa & = \sum_{j} \sin^{2}\omega t_{j}, \\ \nonumber
\CS & = \sum_{j} \cos\omega t_{j} \sin\omega t_{j},
\end{align}
can be evaluated in a single run through the time series, giving a
total of $\Nfreq(2 \Nt + 3)$ trig evaluations for the full periodogram
--- a factor $\sim 2$ improvement.

\section{\culsp: a GPU Lomb-Scargle Periodogram
  Code} \label{sec:culsp}

\begin{figure*}[ht]
\hspace*{0.04\textwidth}
\begin{minipage}[t]{0.45\textwidth}
\begin{lstlisting}[name=source,basicstyle=\scriptsize\tt,escapechar=@,frame=single]
__global__ void
culsp_kernel(float *d_t, float *d_X, float *d_P, @\label{src:args-s}@
             float df, int N_t) @\label{src:args-e}@
{

  __shared__ float s_t[BLOCK_SIZE];
  __shared__ float s_X[BLOCK_SIZE];

  // Calculate the frequency

  float f = (blockIdx.x*BLOCK_SIZE+threadIdx.x+1)*df; @\label{src:freq}@

  // Calculate the various sums

  float XC = 0.f;
  float XS = 0.f;
  float CC = 0.f;
  float CS = 0.f;

  float XC_chunk = 0.f;
  float XS_chunk = 0.f;
  float CC_chunk = 0.f;
  float CS_chunk = 0.f;

  int j;

  for(j = 0; j < N_t; j += BLOCK_SIZE) { @\label{src:outerloop-s}@

    // Load the chunk into shared memory

    __syncthreads(); @\label{src:load-s}@

    s_t[threadIdx.x] = d_t[j+threadIdx.x];  
    s_X[threadIdx.x] = d_X[j+threadIdx.x];  

    __syncthreads(); @\label{src:load-e}@

    // Update the sums

    #pragma unroll @\label{src:unroll}@
    for(int k = 0; k < BLOCK_SIZE; k++) { @\label{src:innerloop-s}@

      // Range reduction

      float ft = f*s_t[k];
      ft -= rintf(ft); @\label{src:argred}@
\end{lstlisting}
\end{minipage}
\hspace*{0.04\textwidth}
\begin{minipage}[t]{0.45\textwidth}
\begin{lstlisting}[name=source,basicstyle=\scriptsize\tt,escapechar=@,frame=single]

      float c;
      float s;

      __sincosf(TWOPI*ft, &s, &c); @\label{src:sincos}@

      XC_chunk += s_X[k]*c;
      XS_chunk += s_X[k]*s;
      CC_chunk += c*c;
      CS_chunk += c*s;

    } @\label{src:innerloop-e}@

    XC += XC_chunk; @\label{src:sum-update-s}@
    XS += XS_chunk;
    CC += CC_chunk;
    CS += CS_chunk; @\label{src:sum-update-e}@

    XC_chunk = 0.f;
    XS_chunk = 0.f;
    CC_chunk = 0.f;
    CS_chunk = 0.f;
    
  } @\label{src:outerloop-e}@

  float SS = (float) N_t - CC; @\label{src:SS}@

  // Calculate the tau terms

  float ct;
  float st;

  __sincosf(0.5f*atan2(2.f*CS, CC-SS), &st, &ct); @\label{src:sin-cos-tau}@

  // Calculate P

  d_P[blockIdx.x*BLOCK_SIZE+threadIdx.x] = @\label{src:P}@ 
      0.5f*((ct*XC + st*XS)*(ct*XC + st*XS)/
	    (ct*ct*CC + 2*ct*st*CS + st*st*SS) + 
	    (ct*XS - st*XC)*(ct*XS - st*XC)/
	    (ct*ct*SS - 2*ct*st*CS + st*st*CC));

  // Finish

}

\end{lstlisting}
\end{minipage}
\caption{Abridged source for the \culsp\ computation
  kernel.} \label{fig:source}
\end{figure*}

\subsection{Overview} \label{ssec:culsp-over}

This section introduces \culsp, a Lomb-Scargle periodogram code
implemented within NVIDIA's CUDA framework. Below, I provide a brief
technical overview of CUDA. Section~\ref{ssec:culsp-design} then
reviews the general design of \culsp, and
Section~\ref{ssec:culsp-source} narrates an abridged version of the
kernel source. The full source, which is freely redistributable under
the GNU General Public License, is provided in the accompanying
on-line materials.

\subsection{The CUDA Framework} \label{ssec:culsp-cuda}

A CUDA-enabled GPU comprises one or more streaming multiprocessors
(SMs), themselves composed of a number\footnote{Eight, for the GPUs
  considered in the present work.} of scalar processors (SPs) that are
functionally equivalent to processor cores. Together, the SPs allow an
SM to support concurrent execution of blocks of up to 512
threads. Each thread applies the same computational kernel to an
element of an input stream. Resources at a thread's disposal include
its own register space; built-in integer indices uniquely identifying
the thread; shared memory accessible by all threads in its parent
block; and global memory accessible by all threads in all
blocks. Reading or writing shared memory is typically as fast as
accessing a register; however, global memory is two orders of
magnitude slower.

CUDA programs are written in the C language with extensions that allow
computational kernels to be defined and launched, and the differing
types of memory be allocated and accessed. A typical program will
transfer input data from CPU memory to GPU memory; launch one or more
kernels to process these data; and then copy the results back from GPU
to CPU. Executables are created using the \texttt{nvcc} compiler from
the CUDA software development kit (SDK).

A CUDA kernel has access to the standard C mathematical functions. In
some cases, two versions are available (`library' and `intrinsic'),
offering different trade-offs between precision and speed \citep[see
Appendix C of][]{NVI2010}. For the sine and cosine functions, the
library versions are accurate to within 2 units of last place, but are
very slow because the range-reduction algorithm --- required to bring
arguments into the $(-\pi/4,\pi/4)$ interval --- spills temporary
variables to global memory. The intrinsic versions do not suffer this
performance penalty, as they are hardware-implemented in two special
function units (SFUs) attached to each SM. However, they become
inaccurate as their arguments depart from the $(-\pi,\pi)$
interval. As discussed below, this inaccuracy can be remedied through
a very simple range-reduction procedure.

\subsection{Code Design} \label{ssec:culsp-design}

The \culsp\ code is a straightforward CUDA implementation of the L-S
periodogram in its refactored form
(equations~\ref{eqn:tau-refac}--\ref{eqn:sums}). A uniform frequency grid
is assumed,
\begin{equation} \label{eqn:freq}
f_{i} = i\,\dfreq \quad (i = 1,\ldots,\Nfreq),
\end{equation}
where the frequency spacing and number of
frequencies are determined from
\begin{equation} \label{eqn:dfreq}
\dfreq = \frac{1}{\Fover(t_{\Nt} - t_{1})}
\end{equation}
and
\begin{equation} \label{eqn:Nfreq} 
\Nfreq = \frac{\Fhigh \Fover \Nt}{2},
\end{equation}
respectively. The user-specified parameters \Fover\ and \Fhigh\
control the oversampling and extent of the periodogram; $\Fover=1$
gives the characteristic sampling established by the length of the
time series, while $\Fhigh=1$ gives a maximum frequency equal to the
mean Nyquist frequency $\freqNy = \Nt/[2 (t_{\Nt} - t_{1})]$.

The input time series is read from disk and pre-processed to have zero
mean and unit variance, before being copied to GPU global
memory. Then, the computational kernel is launched for \Nfreq\ threads
arranged into blocks of size \Nblock\footnote{Set to 256 throughout
  the present work; tests indicate that larger or smaller values give
  a slightly reduced performance.}; each thread handles the
periodogram calculation at a single frequency. Once all calculations
are complete, the periodogram is copied back to CPU memory, and from
there written to disk.

The sums in equation~(\ref{eqn:sums}) involve the entire time series. To
avoid a potential memory-access bottleneck, and to improve accuracy,
\culsp\ partitions these sums into chunks equal in size to the thread
block size \Nblock. The time-series data required to evaluate the sums
for a given chunk are copied from (slow) global memory into (fast)
shared memory, with each thread in a block transferring a single
$(t_{j},X_{j})$ pair. Then, all threads in the block enjoy fast access
to these data when evaluating their respective per-chunk sums.

\subsection{Kernel Source} \label{ssec:culsp-source}

Figure~\ref{fig:source} lists abridged source for the \culsp\
computational kernel. This is based on the full version supplied in
the on-line materials, but special-case code (handling situations
where \Nt\ is not an integer multiple of \Nblock) has been removed
to facilitate the discussion.

The kernel accepts five arguments
(lines~\ref{src:args-s}--\ref{src:args-e} of the listing). The first
three are array pointers giving the global-memory addresses of the
time-series (\lstinline|d_time| and \lstinline|d_data|) and the output
periodogram (\lstinline|d_P|). The remaining two give the frequency
spacing of the periodogram (\lstinline|df|) and the number of points
in the time series (\lstinline|N_t|). The former is used on
line~\ref{src:freq} to evaluate the frequency from the thread and
block indices; the macro \lstinline|BLOCK_SIZE| is expanded by the
pre-processor to the thread block size \Nblock.

Lines \ref{src:outerloop-s}--\ref{src:outerloop-e} construct the sums
of equation~(\ref{eqn:sums}), following the chunk partitioning
approach described above (note, however, that the \lstinline|SS| sum
is not calculated explicitly, but reconstructed from \lstinline|CC| on
line~\ref{src:SS}). Lines~\ref{src:load-s}--\ref{src:load-e} are
responsible for copying the time-series data for a chunk from global
memory to shared memory; the \lstinline|__syncthreads()| instructions
force synchronization across the whole thread block, to avoid
potential race conditions. The inner loop
(lines~\ref{src:innerloop-s}--\ref{src:innerloop-e}) then evaluates
the per-chunk sums; the \lstinline|\#pragma unroll| directive on
line~\ref{src:unroll} instructs the compiler to completely unroll this
loop, conferring a significant performance increase.

The sine and cosine terms in the sums are evaluated simultaneously
with a call to CUDA's intrinsic \lstinline|__sincosf()| function
(line~\ref{src:sincos}). To maintain accuracy, a simple range
reduction is applied to the phase \lstinline|ft| by subtracting the
nearest integer [as calculated using \lstinline|rintf()|;
line~\ref{src:argred}]. This brings the argument of
\lstinline|__sincosf()| into the interval $(-\pi,\pi)$, where its
maximum absolute error is $2^{-21.41}$ for sine and $2^{-21.19}$ for
cosine \citep[see Table C-3 of][]{NVI2010}.

\section{Benchmarking Calculations} \label{sec:calc}

\subsection{Test Configurations} \label{ssec:calc-config}

\begin{deluxetable}{lcccc}
\tablecaption{Specifications for the two GPUs used in the benchmarking.\label{tab:gpus}}
\tablehead{
GPU & SMs & SPs & Clock
(GHz) &
Memory (MB)
}
\tablecolumns{4}
\tabletypesize{\footnotesize}
\startdata
GeForce 8400 GS & 1 & 8 & 1.4 &512 \\
Tesla C1060 & 30 & 240 & 1.3 & 4096
\enddata
\end{deluxetable}


This section compares the accuracy and performance of \culsp\ against
an equivalent CPU-based code. The test platform is a Dell Precision
490 workstation, containing two Intel 2.33 GHz Xeon E5345 quad-core
processors and 8 GB of RAM. The workstation also hosts a pair of
NVIDIA GPUs: a Tesla C1060 populating the single PCI Express (PCIe)
$\times16$ slot, and a GeForce 8400 GS in the single legacy PCI
slot. These devices are broadly representative of the opposite ends of
the GPU market.  The 8400 GS is an entry-level product based on the
older G80 hardware architecture (the first to support CUDA), and
contains only a single SM. The C1060 is built on the newer GT200
architecture (released 2008/2009), and with 30 SMs represents one of
the most powerful GPUs in NVIDIA's portfolio. The technical
specifications of each GPU are summarized in Table~\ref{tab:gpus}.

The CPU code used for comparison is \lsp, a straightforward port of
\culsp\ to ISO C99 with a few modifications for performance and
language compliance. The sine and cosine terms are calculated via
separate calls to the \lstinline|sinf()| and \lstinline|cosf()|
functions, since there is no \lstinline|sincosf()| function in
standard C99. The argument reduction step uses an integer cast instead
of \lstinline|rintf()|; this allows the compiler to vectorize the
inner loops, greatly improving performance while having a negligible
impact on results. Finally, the outer loop over frequency is trivially
parallelized using an OpenMP directive, so that all available CPU
cores can be utilized. Source for \lsp\ is provided in the
accompanying on-line materials.

The Precision 490 workstation runs 64-bit Gentoo Linux. GPU
executables are created with the 3.1 release of the CUDA SDK, which
relies on GNU \texttt{gcc} 4.4 as the host-side compiler. CPU
executables are created with Intel's \texttt{icc} 11.1 compiler, using
the \texttt{-O3} and \texttt{-xHost} optimization flags.

\subsection{Accuracy} \label{ssec:calc-acc}

\begin{figure}[htb!]
\epsscale{1}
\begin{centering}
\includegraphics{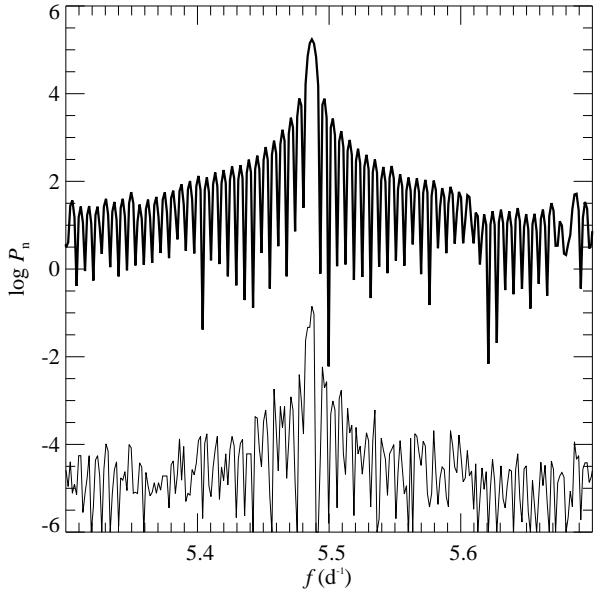}
\caption{Part the L-S periodogram for \VAql, evaluated using the \lsp\
  code (thick curve). The thin curve shows the absolute deviation
  $|\Pnormc-\Pnorml|$ of the corresponding periodogram evaluated using
  the \culsp\ code. The strong peak corresponds to the star's dominant
  0.18-d pulsation mode.} \label{fig:pgram}
\end{centering}
\end{figure}

\begin{figure}[htb!]
\epsscale{1}
\begin{centering}
\includegraphics{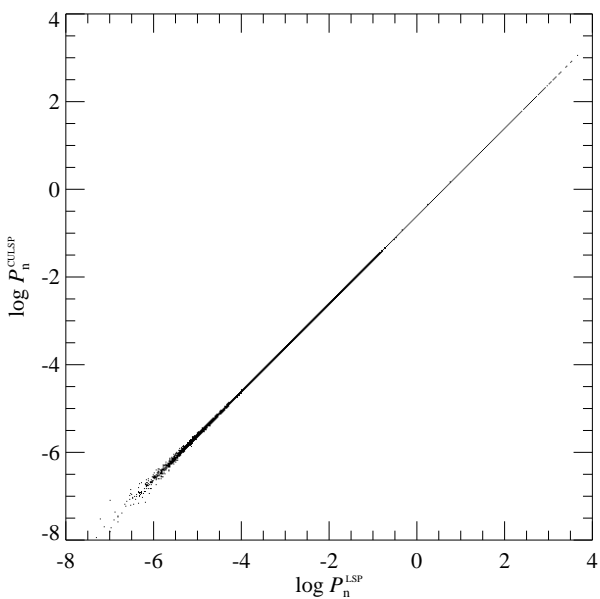}
\caption{A scatter plot of the L-S periodogram for \VAql, evaluated
  using the \lsp\ (abscissa) and \culsp\ (ordinate)
  codes.} \label{fig:accuracy}
\end{centering}
\end{figure}

As the validation dataset for comparing the accuracy of \culsp\ and
\lsp, I use the 150-day photometric time series of the $\beta$ Cephei
pulsator \VAql\ (HD~180642) obtained by the \emph{CoRoT} mission
\citep{Bel2009}. The observations comprise 382,003 flux measurements
(after removal of points flagged as bad), sampled unevenly (in the
heliocentric frame) with an average separation of 32\,s.

Figure~\ref{fig:pgram} plots the periodogram of \VAql\ evaluated using
\lsp, over a frequency interval spanning the star's dominant 0.18\,d
pulsation mode \citep[see][]{Wae1998}. Also shown in the figure is the
absolute deviation $|\Pnormc-\Pnorml|$ of the corresponding
periodogram evaluated using \culsp\ (running on either GPU --- the
results are identical). The figure confirms that, at least over this
particular frequency interval, the two codes are in good agreement
with one another; the relative error is on the order of $10^{-6}$.

To explore accuracy over the full frequency range,
Fig.~\ref{fig:accuracy} shows a scatter plot of \Pnorml\ against
\Pnormc. Very few of the $\Nfreq = 1,528,064$ points in this plot
depart to any significant degree from the diagonal line $\Pnorml =
\Pnormc$. Those that do are clustered in the $\Pnorm \ll 1$ corner of
the plot, and are therefore associated with the noise in the light
curve rather than any periodic signal. Moreover, the maximum absolute
difference in the periodogram FAPs (equation~\ref{eqn:fap}) across all
frequencies is $4.1 \times 10^{-5}$, which is negligible.

\subsection{Performance} \label{ssec:calc-perf}

\begin{figure}[htb!]
\epsscale{1}
\begin{centering}
\includegraphics{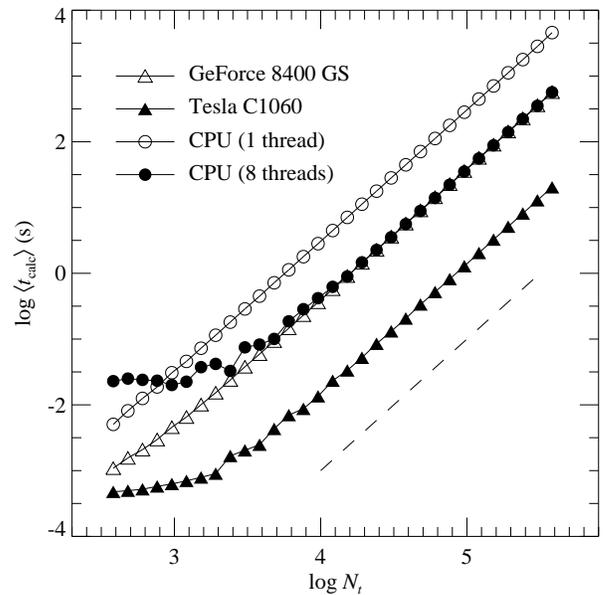}
\caption{Mean calculation times \tcalcm\ for the L-S periodogram,
  evaluated using the \culsp\ (triangles) and \lsp\ (circles)
  codes. The dashed line, with slope $\diff\log\tcalcm/\diff\log\Nt =
  2$, indicates the asymptotic scaling of the periodogram
  algorithm.} \label{fig:timing}
\end{centering}
\end{figure}

\begin{deluxetable}{llrr}
\tablecaption{Periodogram calculation times.\label{tab:timing}}
\tablehead{
Code & Platform & \tcalcm\ (s) & \tcalcs\ (s) }
\tablecolumns{4}
\tabletypesize{\footnotesize}
\startdata
\culsp & GeForce 8400 GS &   570   & 0.0093  \\
\culsp & Tesla C1060     &    20.3 & 0.00024 \\
\lsp   & CPU (1 thread)  &  4570   & 14      \\
\lsp   & CPU (8 threads) &   566   & 6.9
\enddata
\end{deluxetable}

Code performance is measured by averaging the \VAql\ periodogram
calculation time \tcalc\ over five executions. These timings exclude
the overheads incurred by disk input/output and from rectifying light
curves to zero mean an unit variance.  Table~\ref{tab:timing} lists
the mean \tcalcm\ and associated standard deviation \tcalcs\ for
\culsp\ running on both GPUs, and for \lsp\ running with a single
OpenMP thread (equivalent to a purely serial CPU implementation), and
with 8 OpenMP threads (one per workstation core).

With just one thread, \lsp\ is significantly outperformed by \culsp\
on either GPU. Scaling up to 8 threads shortens the calculation time
by a factor $\sim 8$, indicating near-ideal parallelization;
nevertheless, the two CPUs working together only just manage to beat
the GeForce 8400 GS, and are still a factor $\sim 28$ slower than the
Tesla C1060. Perhaps surprisingly, the latter ratio is \emph{greater}
than suggested by comparing the theoretical peak floating-point
performance of the two platforms --- 74.6 GFLOPS (billions of
floating-point operations per second) for all 8 CPU cores, versus 936
GFLOPS for the C1060. This clearly warrants further investigation.

Profiling with the GNU \texttt{gprof} tool indicates that the major
bottleneck in \lsp, accounting for 80\% of \tcalcm, is the
\lstinline|__svml_sincosf4()| function from Intel's Short Vector Math
Library. This function evaluates four sine/cosine pairs at once by
leveraging the SSE2 instructions of modern x86-architecture CPUs.
Microbenchmarking reveals that a \lstinline|__svml_sincosf4()| call
costs $\sim 45.6$ clock cycles, or $\sim 11.4$ cycles per sine/cosine
pair. In contrast, thanks to its two special function units, a GPU SM
can evaluate a sine/cosine pair in a single cycle \citep[see Appendix
G.3.1 of][]{NVI2010}. Scaling these values by the appropriate clock
frequencies and processor counts, the sine/cosine throughput on all 8
CPU cores is 1.6 billion operations per second (GOPS), whereas on the
30 SMs of the C1060 it is 39 GOPS, around 23 times faster. Of course,
it should be recalled that the GPU \lstinline|__sincosf()| function
operates at a somewhat-reduced precision (see
Section~\ref{ssec:culsp-source}). In principle, the CPU throughput
could be improved by developing a similar reduced-precision function
to replace \lstinline|__svml_sincosf4()|. However, it seems unlikely
that a \emph{software} routine could ever approach the throughput of
the dedicated hardware in the SFUs.

The disparity between sine/cosine throughput accounts for most of the
factor $\sim 28$ performance difference between \culsp\ and \lsp,
noted above. The remainder comes from the ability of an SM to execute
instructions simultaneously on its SFUs and scalar processors. That
is, the sine/cosine evaluations can be undertaken \emph{in parallel}
with the other arithmetic operations involved in the periodogram
calculation.

Looking now at the memory performance of \culsp, NVIDIA's
\texttt{cudaprof} profiling tool indicates that almost all reads from
global memory are coalesced, and that no bank conflicts arise when
reading from shared memory. Thus, the GPU memory accesses patterns can
be considered close to optimal. The combined time spent copying data
from CPU to GPU and vice versa is $\sim 6\,{\rm ms}$ on the C1060, and
$\sim 29\,{\rm ms}$ on the 8400 GS; while these values clearly reflect
the bandwidth difference between the PCIe and PCI slots hosting the
GPUs, neither makes any appreciable contribution to the execution
times listed in Table~\ref{tab:timing}.

To round off the present discussion, I explore how \culsp\ and \lsp\
perform with different-sized datasets. The analysis in
Section~\ref{sec:l-s} indicates a periodogram workload scaling as
$\mathcal{O}(\Nfreq \Nt)$. With the number of frequencies following
$\Nfreq \propto \Nt$ (equation~\ref{eqn:Nfreq}), \tcalc\ should
therefore scale proportionally with $\Nt^{2}$ --- as in fact already
claimed in Introduction. To test this expectation,
Fig.~\ref{fig:timing} shows a log-log plot of \tcalcm\ as a function
of \Nt, for the same configurations as in Table~\ref{tab:timing}. The
light curve for a given \Nt\ is generated from the full \VAql\ light
curve by uniform down-sampling.

In the limit of large \Nt, all curves asymptote toward a slope
$\diff\log\tcalcm/\diff\log\Nt = 2$, confirming the hypothesized
$\Nt^{2}$ scaling. At smaller \Nt, departures from this scaling arise
from computational overheads that are not directly associated with the
actual periodogram calculation. These are most clearly seen in the
\lsp\ curve for 8 threads, which approaches a constant $\log \tcalcm
\approx -1.5$ independent of \Nt\ --- perhaps due to memory cache
contention between the different threads.

\section{Discussion} \label{sec:discuss}

\subsection{Cost/Benefit Analysis} \label{ssec:discuss-cost}

To establish a practical context for the results of the preceding
sections, I briefly examine the price vs. performance of the CPU and
GPU platforms. At the time of writing, the manufacturer's bulk
(1,000-unit) pricing for a pair of Xeon E5345 CPUs is $2\times\$455$,
while a Tesla C1060 retails for around \$1,300 and a GeForce 8400 GS
for around \$50. First considering the C1060, the 50\% greater cost of
this device (relative to the CPUs) brings almost a factor thirty
reduction in periodogram calculation time --- an impressive degree of
leveraging. However, its hefty price tag together with demanding
infrastructure requirements (dedicated PCIe power connectors,
supplying up to 200\,W), means that it may not be the ideal GPU choice
in all situations.

The 8400 GS offers a similar return-on-investment at a much-more
affordable price: almost the same performance as the two CPUs at
one-twentieth of the cost. This heralds the possibility of budget GPU
computing, where a low-end desktop computer is paired with an
entry-level GPU, to give performance exceeding high-end, multi-core
workstations for a price tag of just a few hundred dollars. Indeed,
many desktop computers today, or even laptops, are already well
equipped to serve in this capacity.

\subsection{Applications} \label{ssec:discuss-app}

An immediate application of \culsp\ is analysis of the photometric
time series obtained by ongoing satellite missions such as \emph{MOST}
\citep{Wal2003}, \emph{CoRoT} \citep{Auv2009}, and \emph{Kepler}
\citep{Koc2010}. These datasets are typically very large ($\Nt \gtrsim
10^{5}$), leading to a significant per-star cost for calculating a
periodogram. When this cost is multiplied by the number of targets
being monitored (in the cast of \emph{Kepler}, again $\gtrsim
10^{5}$), the overall computational burden becomes very steep. Looking
into the near future, similar issues will be faced with ground-based
time-domain facilities such as Pan-STARRS \citep{Kai2002} and the
Large Synoptic Survey Telescope \citep{LSST2009}. It is hoped that
\culsp, or an extension to other related periodograms (see below),
will help resolve these challenges.

An additional application of \culsp\ is in the interpretation of
periodograms. Equation~(\ref{eqn:fap}) presumes that the \Pnorm\ at
each frequency in the periodogram is independent of the others. This
is not necessarily the case, and the exponent in the equation should
formally be replaced by some number \Nfreqind\ representing the number
of independent frequencies. \citet{HorBal1986} pioneered the use of
simulations to estimate \Nfreqind\ empirically, and similar
Monte-Carlo techniques have since been applied to explore the
statistical properties of the L-S periodogram in detail
\citep[see][and references therein]{Fre2008}. The bottleneck in these
simulations are the many periodogram evaluations, making them strong
candidates for GPU acceleration.

\subsection{Future Work} \label{ssec:discuss-future}

Recognizing the drawbacks of being wedded to one particular
hardware/software vendor, a strategically important future project
will be to port \culsp\ to Open CL (Open Computing Language) --- a
recently developed standard for programming devices such as multi-core
CPUs and GPUs in a platform-neutral manner \citep[see,
e.g.,][]{Sto2010}. There is also considerable scope for applying the
lessons learned herein to other spectral analysis
techniques. \citet{Shr2001} and \citet{ZecKur2009} generalize the L-S
periodogram to allow for the fact that the time-series mean is
typically not known \emph{a priori}, but instead estimated from the
data themselves. The expressions derived by these authors involve sums
having very similar forms to equation~(\ref{eqn:sums}); thus, it
should prove trivial to develop GPU implementations of the generalized
periodograms. The multi-harmonic periodogram of \citet{Sch1996} and
the SigSpec method of \citet{Ree2007} also appear promising candidates
for implementation on GPUs, although algorithmically they are rather
more-complex.

Looking at the bigger picture, while the astronomical theory and
modeling communities have been quick to recognize the usefulness of
GPUs (see Section~\ref{sec:intro}), progress has been more gradual in
the observational community; radio correlation is the only significant
application to date \citep{Way2009}. It is my hope that the present
paper will help illustrate the powerful data-analysis capabilities of
GPUs, and demonstrate strategies for using these devices effectively.


\acknowledgments

I thank Dr. Gordon Freeman for the initial inspiration to explore this
line of research, and the anonymous referee for many helpful
suggestions that improved the manuscript. I moreover acknowledge
support from NSF \emph{Advanced Technology and Instrumentation} grant
AST-0904607. The Tesla C1060 GPU used in this study was donated by
NVIDIA through their Professor Partnership Program, and I have made
extensive use of NASA's Astrophysics Data System bibliographic
services.


\bibliographystyle{aastex}
\bibliography{gpu-period}

\end{document}